\documentclass[a4 paper, 12 pt] {article}

\usepackage{graphics, graphicx}

\begin{document}

\title{\bf{Knots, Braids and Hedgehogs from the Eikonal
Equation}}
\author{A. Wereszczy\'{n}ski \thanks{wereszcz@alphas.if.uj.edu.pl}
       \\
       \\ Institute of Physics,  Jagiellonian University,
       \\ Reymonta 4, Krak\'{o}w, Poland}
\maketitle
\begin{abstract}
The complex eikonal equation in the three space dimensions is
considered. We show that apart from the recently found torus knots
this equation can also generate other topological configurations
with a non-trivial value of the $\pi_2(S^2)$ index: braided open
strings as well as hedgehogs. In particular, cylindric strings
i.e. string solutions located on a cylinder with a constant radius
are found. Moreover, solutions describing strings lying on an
arbitrary surface topologically equivalent to cylinder are
presented. We discus them in the context of the eikonal knots. The
physical importance of the results originates in the fact that the
eikonal knots have been recently used to approximate the
Faddeev-Niemi hopfions.
\end{abstract}
\newpage
%%%%%%%%%%%%%%%%%%%%%%%%%%%%%%%%%%%%%%%%%%%%%%%%%%%%%%%%%%%%%%%
\section{\bf{Introduction}}
%%%%%%%%%%%%%%%%%%%%%%%%%%%%%%%%%%%%%%%%%%%%%%%%%%%%%%%%%%%%%%%
It has been recently proved that the complex eikonal equation
\begin{equation}
(\nabla u )^2 =0, \label{eikonal}
\end{equation}
where $u$ is a complex scalar field plays a prominent role in high
dimensional soliton theories. It had been recently shown that it
is an integrability condition for a very wide class of $CP^N$
theories in (2+1) and (3+1) dimensional space-time
\cite{ferreira}, \cite{ferreira2}, \cite{adamnew} (for further
results in any dimension and for other models see \cite{ferreira},
\cite{high}). Indeed, when one assumes that the scalar field must
fulfill (apart from the pertinent dynamical equations of motion)
this constrain, then an integrable submodel can be defined. Here,
integrability is understood as existence of an infinite family of
local conserved currents. On the other hand, it is a conventional
wisdom known form soliton theories in (1+1) dimensions that the
appearance of an infinite family of the conserved currents is
closely related with the existence of topological solitons. The
reason is simple - solitons are observed in systems with highly
non-trivial relations between degrees of freedom, what is
equivalent to high degree of symmetries in the system. Usually,
such symmetries result in appearance of the conserved currents.
\\
In fact, it has been checked by analytical and numerical
calculations that spectrum of the solutions of such integrable
submodels can contain solitons \cite{aratyn1}. Let us only mention
the Nicole model \cite{nicole}, where hopfion i.e. a soliton with
a non-zero value of the Hopf index has been found.
\\
This powerful approach has been recently applied also to the
Faddeev-Niemi model \cite{niemi1}. It is believed that this model,
based on an unit, three component vector field $\vec{n}$ living in
(3+1) Minkowski space-time can be a good candidate for the
effective model for the low energy quantum gluodynamics
\cite{niemi1}, \cite{cho}, \cite{shabanov}, \cite{pak}. Here,
effective particles i.e. glueballs are described as knotted
configurations of the gauge field \cite{niemi2} \footnote{Knots
found application also in other branches of physics \cite{physics}
as well as chemistry \cite{chem} and biology \cite{biol}.}.
Indeed, such knotted solitons have been obtained in numerical work
\cite{helmut}, \cite{battyde}, \cite{salo}. In the case of the
Faddeev-Niemi model, the construction of the integrable submodel
can be easily done \cite{aratyn1} after taking into account the
standard stereographic projection, which relates the original
vector field with the complex scalar field
\begin{equation}
\vec{n}= \frac{1}{1+|u|^2} ( u+u^*, -i(u-u^*), |u|^2-1).
\label{stereograf}
\end{equation}
Unfortunately, even the submodel is too complicated and no exact,
knotted solutions have been reported. In spite of that, the
eikonal equation appears to be very helpful in the construction of
an approximation to the Faddeev-Niemi hopfions \cite{ja1}. It
provides a framework which enable us to analytically investigate
qualitative (topology, shape) as well as quantitative (energy)
features of the Faddeev-Niemi solitons. For example, using knotted
solutions of the eikonal equation, so-called eikonal knots
\cite{adam}, approximated but analytical solutions of the
Faddeev-Niemi model have been constructed. These solutions can
represent the numerically obtained Faddeev-Niemi hopfions with
approximately $20\%$ accuracy \cite{ja1}. Additionally, it was
observed that energy of the eikonal knots (calculated in the
Faddeev-Niemi model) with a fixed topological charge is bounded
from below. All these facts might suggest that the topological
content of the Faddeev-Niemi model and the eikonal equation is
similar.
\\
One can also notice that the eikonal knots , unlikely hopfions
obtained in the dynamical toy-models \cite{aratyn2}, \cite{ja2},
\cite{ferreira3}, are really knotted objects. Because of that they
can be applied in other physical or biological contexts as well.
\\
Nonetheless, due to the complicatedness of the toroidal symmetry
many properties of the eikonal knots and, in consequence, the
Faddeev-Niemi hopfions have not been understood sufficiently.
\\
First of all, the existence of non-torus knots, that is knots
which cannot be plotted on a torus (for instance the figure-eight
knot) is still an open question. Secondly, it is also unknown
whether equation (\ref{eikonal}) admits knots which are located on
arbitrary closed surfaces topologically equivalent to torus. These
'deformed knots' might be very helpful in obtaining less energy
approximation to the Faddeev-Niemi hopfions.
\\
Moreover, since the vector field $\vec{n}$ can, in principle, form
other topological objects (referred as monopoles and open
strings), it would be important to know what kinds of topological
structures are admitted by the eikonal equation.
\\
In the present paper we would like to face some of the upper
mentioned problems. It is done in the language of braided string
solutions of the eikonal equations, which are found in the next
section. Then, taking advantage of the relationship between braids
(open strings) and knots (closed string), we can extend our
results on the eikonal knots.
%%%%%%%%%%%%%%%%%%%%%%%%%%%%%%%%%%%%%%%%%%%%%%%%%%%%%%%%%%%%%%%
\section{\bf{Topological Strings}}
%%%%%%%%%%%%%%%%%%%%%%%%%%%%%%%%%%%%%%%%%%%%%%%%%%%%%%%%%%%%%%%
In order to deal with strings it is natural to introduce the
cylindrical coordinates. Then the eikonal equation is given as
follows
\begin{equation}
 (\partial_{\rho} u)^2 + \frac{1}{\rho^2}(\partial_{\phi} u)^2
+(\partial_{z} u)^2=0, \label{eikonal walc}
\end{equation}
where the solution is assumed to have the following form
\begin{equation}
u(\rho,\phi,z)=R(\rho)\Phi(\phi)Z(z). \label{sep walc}
\end{equation}
Using this Ansatz one can rewrite the eikonal equation and find
\begin{equation}
\frac{1}{R}(\partial_{\rho} R)^2 + \frac{1}{\rho^2
\Phi}(\partial_{\phi} \Phi)^2 +\frac{1}{Z}(\partial_{z} Z)^2=0.
\end{equation}
This equation can be expressed in terms of three first order
ordinary differential equations. Let us start with the equation
for the $z$ variable. Namely,
\begin{equation}
Z'^2_{z}(z)+k^2Z^2(z)=0, \label{eq z}
\end{equation}
where $k^2$ is a positive constant. Thus
\begin{equation}
Z(z)=Ae^{\pm ikz}, \label{sol z}
\end{equation}
where $A$ is a complex constant. The pertinent equation for the
angular coordinate takes this form
\begin{equation}
\Phi'^2_{\phi }(\phi) +n^2 \Phi (\phi)=0 \label{eq phi}
\end{equation}
and possesses solutions
\begin{equation}
\Phi(\phi)=Be^{\pm i n\phi}, \label{sol phi}
\end{equation}
where $B$ is a complex constant. To guarantee the uniqueness of
the solution, the parameter $n$ has to be an integer number.
Finally, taking into account (\ref{eq z}) and (\ref{eq phi}) we
obtain the equation for $\rho$
\begin{equation}
R'^2_{\rho} - \left( \frac{n^2}{\rho^2} +k^2 \right) R^2 =0.
\label{eq rho}
\end{equation}
It can be solved and we find that
\begin{equation}
R(\rho)= C  \left( \frac{\rho}{n+\sqrt{k^2\rho^2+n^2}}
\right)^{\pm n} e^{\pm \sqrt{k^2\rho^2+n^2}}. \label{sol rho}
\end{equation}
Here $C$ in a complex constant. Now, inserting found solutions
(\ref{sol z}), (\ref{sol phi}) and (\ref{sol rho}) into (\ref{sep
walc}) one can derive a solution of the eikonal equation
\begin{equation}
u(\rho,\phi,z)=C  \left( \frac{\rho}{n+\sqrt{k^2\rho^2+n^2}}
\right)^{ \pm n} e^{\pm \sqrt{k^2\rho^2+n^2}} e^{\pm i (n\phi
+kz)}, \label{anzatz walc}
\end{equation}
with a new constant $C$. Moreover, one can observe that by means
of (\ref{anzatz walc}) we are able to construct more general
solution. Such a solution is just a sum of (\ref{anzatz walc})
solutions
\begin{equation}
u(\rho,\phi,z)=\sum_{j=1}^N C_j  \left(
\frac{\rho}{n_j+\sqrt{k^2_j\rho^2+n^2_j}} \right)^{ \pm n_j} e^{
\pm \sqrt{ k^2_j\rho^2+n^2_j}} e^{ \pm i (n_j\phi
+k_jz)}+c\label{anzatz walc gen}
\end{equation}
with the parameters $k_j$ and $n_j$ obeying the condition
\begin{equation}
\frac{k_j}{n_j} =\mbox{const}.  \label{cond walc}
\end{equation}
Here, $c$ is a new complex constant. One can recognize in this
formula an equivalent of the constant value of the ratio between
'winding' numbers in $\xi$ and $\phi$ directions for the knotted
configurations \cite{ja1}. In both cases this condition guarantees
that only regular, non-intersecting objects can be constructed. No
singular configurations, neither knots nor strings, are allowed by
the eikonal equation.
\\
Of course, upper obtained configuration (\ref{anzatz walc gen}) is
topological non-trivial only if the pertinent topological index
does not vanish. This calculation cab be performed taking into
account the well-known fact that the topological charge of any
rational map of the form $R(z)=p(z)/q(z)$, where $p$ and $q$ are
polynomials without a common divisor, is equal to maximum of the
degree of this polynomial. In our case, for fixed value of the $z$
coordinate, the Ansatz is nothing else but a modified rational
map. Thus, one can check that the standard winding number reads
\begin{equation}
Q=\mbox{max} \{ n_j, j=1..N \}. \label{charge}
\end{equation}
Once again the analogy to the knots is clearly visible. The global
topology is fixed by the part of solution (\ref{anzatz walc gen})
with the biggest $n_j$ whereas local topological structure i.e. a
particular distribution of the topological charge in the
space depends on all $n_j$ (see \cite{ja1}).
\\
Let us now make a digression and notice that our solution
(\ref{anzatz walc}) is also a solution of the massive eikonal
equation
\begin{equation}
(\nabla u )^2 =m^2 u^2, \label{mass eikonal}
\end{equation}
$m^2 >0$, if we substitute $k^2 \rightarrow k^2+m^2$. Then
\begin{equation}
u(\rho,\phi,z)= C  \left(
\frac{\rho}{n+\sqrt{(k^2+m^2)\rho^2+n^2}} \right)^{\pm n} e^{\pm
\sqrt{(k^2+m^2)\rho^2+n^2}} e^{ \pm i (n\phi +kz)}.  \label{sol
rho mass}
\end{equation}
It is straightforward to see that (unlikely the simple massless
eikonal equation) a sum of such solutions does not fulfill the
massive equation.
\\
This solution can be immediately adopted in the case of the
massive eikonal equation in $(2+1)$ dimensions, where we find that
\begin{equation}
u(\rho,\phi)= \left( \frac{\rho}{n+\sqrt{m^2\rho^2+n^2}}
\right)^{\pm n} e^{\pm \sqrt{m^2\rho^2+n^2}} e^{ \pm i n\phi}.
\label{sol rho mass 2d}
\end{equation}
This might be interesting in the context of the non-linear sigma
model in $(2+1)$ dimensions and baby Skyrme models
\cite{baby_skyrme}, \cite{stern} where solutions of the (massless)
eikonal equation are used to build approximated solutions. Thus
all baby skyrmions are long-range objects with power-like
asymptotic behavior. Topological defects which emerge from the
massive eikonal equation are much more well-localized. The unit
vector field fades exponentially as it is expected for massive
fields. In fact, this feature allows us to think about equation
(\ref{mass eikonal}) as a massive counterpart of the standard
eikonal equation. Quite interesting, the massive solution can be
achieved also from a dynamical equation. Namely,
\begin{equation}
\nabla^2 u= m^2_{eff} u, \label{dyn eqmot}
\end{equation}
where $m_{eff}$ denotes an effective mass given as
\begin{equation}
m^2_{eff}=m^2 \left( 1+\frac{1}{\sqrt{m^2r^2+n^2}} \right).
\label{mass}
\end{equation}
We see that for all obtained solutions (massive and massless) the
vector field tends to a constant value at the spatial infinity
$\vec{n} \rightarrow \vec{n}_{\infty}$. Thus the position of the
string is defined as a curve where the vector field points in the
opposite direction i.e.
\begin{equation}
\vec{n}_0=-\vec{n}^{\infty}.  \label{def pos1}
\end{equation}
Solution (\ref{anzatz walc gen}) gives
\begin{equation}
\vec{n}^{\infty}=\lim_{\rho \rightarrow \infty} \vec{n}=(0,0,1),
\label{def pos2}
\end{equation}
where the solution with $(+)$ sing has been chosen. The position
of the string is given by the formula
\begin{equation}
|u|^2=0. \label{def pos3}
\end{equation}
Inserting solution (\ref{anzatz walc gen}) one can find that it is
equivalent to the following algebraic equation
\begin{equation}
\sum_{i,j=1}^N R_{i}R_{j} \cos[ (k_i-k_j)z +(n_i-n_j)\phi ] +
2c_0\sum_{i=1}^N R_i \cos[k_i z +n_i \phi ]+c_0^2=0.
\label{position_solit_gen}
\end{equation}
For simplicity reasons we put $C_i=1$ and $c=c_0 \in R$. In the
next subsections this equation is solved in the simplest but
generic cases.
\begin{figure} \center \resizebox{.1\textwidth}{!}
{\includegraphics{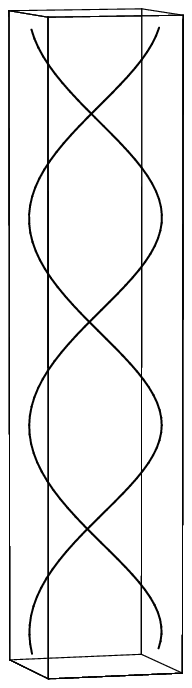}} \hspace{2cm}
\resizebox{.25\textwidth}{!}{\includegraphics{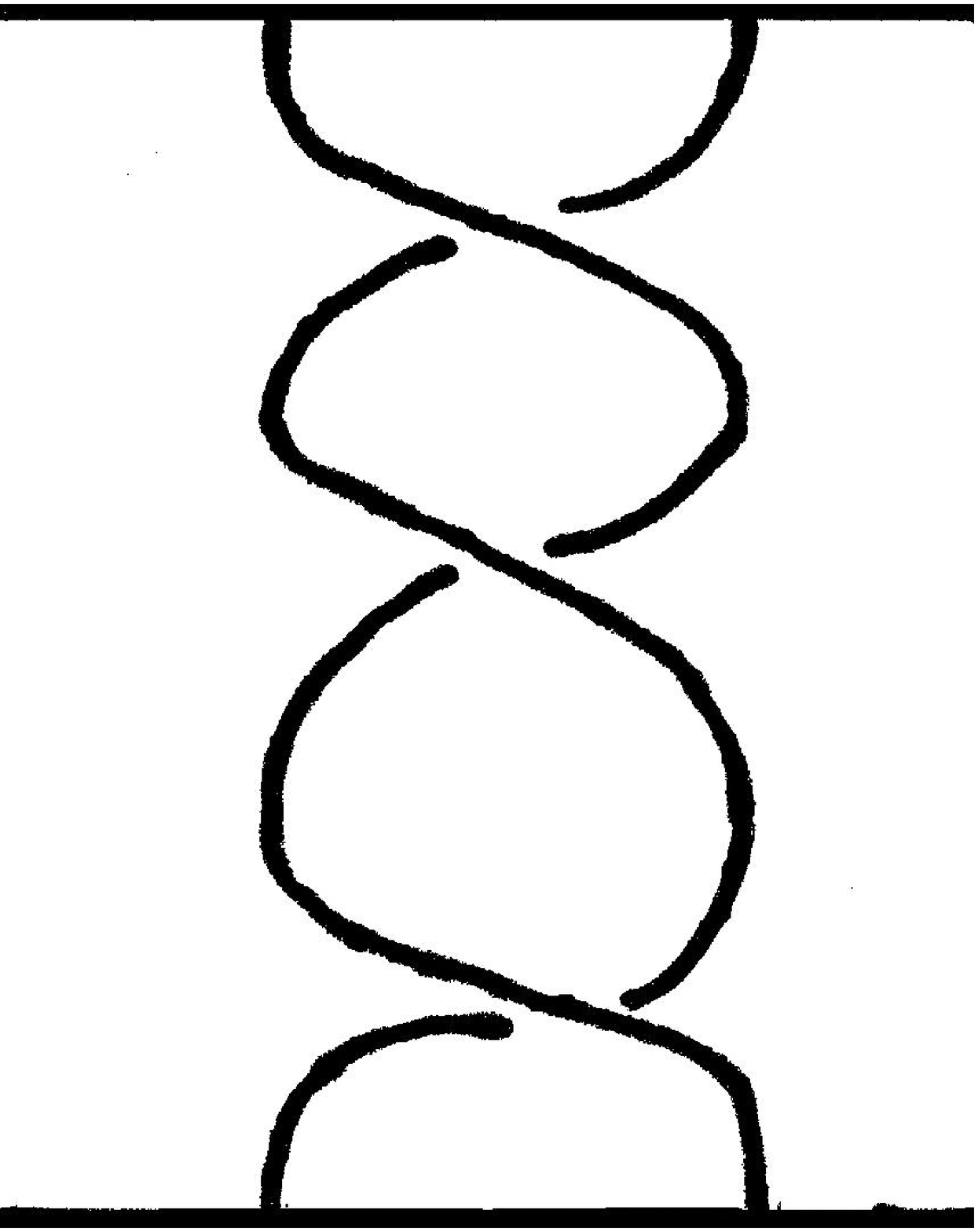}}
\caption{String solution with $n=2$, $k=1$ and schematic two-braid
representation of the trefoil knot.}
\end{figure}
%%%%%%%%%%%%%%%%%%%%%%%%%%%%%%%%%%%%%%%%%%%%%%%%%%%%
\subsection{N=1 case}
%%%%%%%%%%%%%%%%%%%%%%%%%%%%%%%%%%%%%%%%%%%%%%%%%%%%
We start with the simplest case and take into consideration only
one-com\-po\-nent solution (\ref{anzatz walc gen}). Then equation
(\ref{position_solit_gen}) takes the form
\begin{equation}
R^2+2c_0 R \cos[kz +n \phi]+c_0^2=0 \label{position_solit_gen_1}
\end{equation}
and can be easily solved. We obtain that the defects are located
in
\begin{equation}
R(\rho_0)=c_0, \; \; \; kz +n \phi =\pi +2\pi l,
\label{pos_sol_curve1}
\end{equation}
where $l=0,1..n-1$. In other words, as we expected, our solutions
appear to be strings located on a cylinder with a constant radius
$\rho=\rho_0$. For any fixed $z=z_0$ the vector field winds $n$
times around the strings in the $z_0$ plane. Simultaneously, these
$n$ strings wrap infinitely many times around the infinite long
cylinder. There are $k$ coils on the interval $\Delta z=2\pi n$.
\\
In Fig. 1 and Fig. 2 we present a part of two- and three-string
solutions with $n=2, k=1$ and $n=3, k=\frac{2}{3}$ respectively.
It is easy to observe that, after identification of the bases of
the cylinder (i.e. upper and lower ends of the strings), both
configurations can be associated with the trefoil knot. One can
apply this procedure to other multi-string configurations and find
various topologically inequivalent knots. This fact can be
understood from the knot theory point of view. Indeed, it is a
well-known result from knot and braid theory that every torus knot
has a representation in terms of a braid diagram, obtained by the
action of the braiding operators on a trivial braid.
\begin{figure} \center \resizebox{.1\textwidth}{!}
{\includegraphics{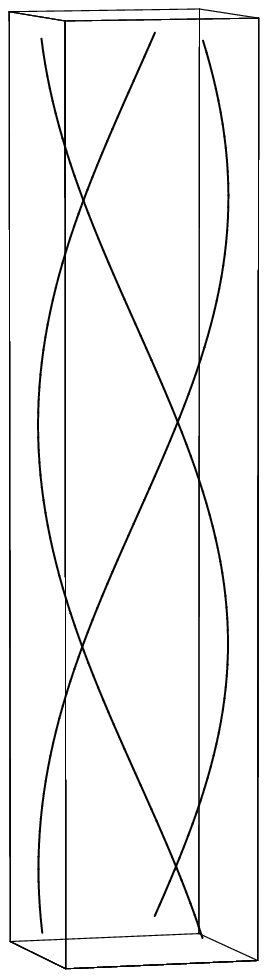}} \hspace{2cm}
\resizebox{.25\textwidth}{!} {\includegraphics{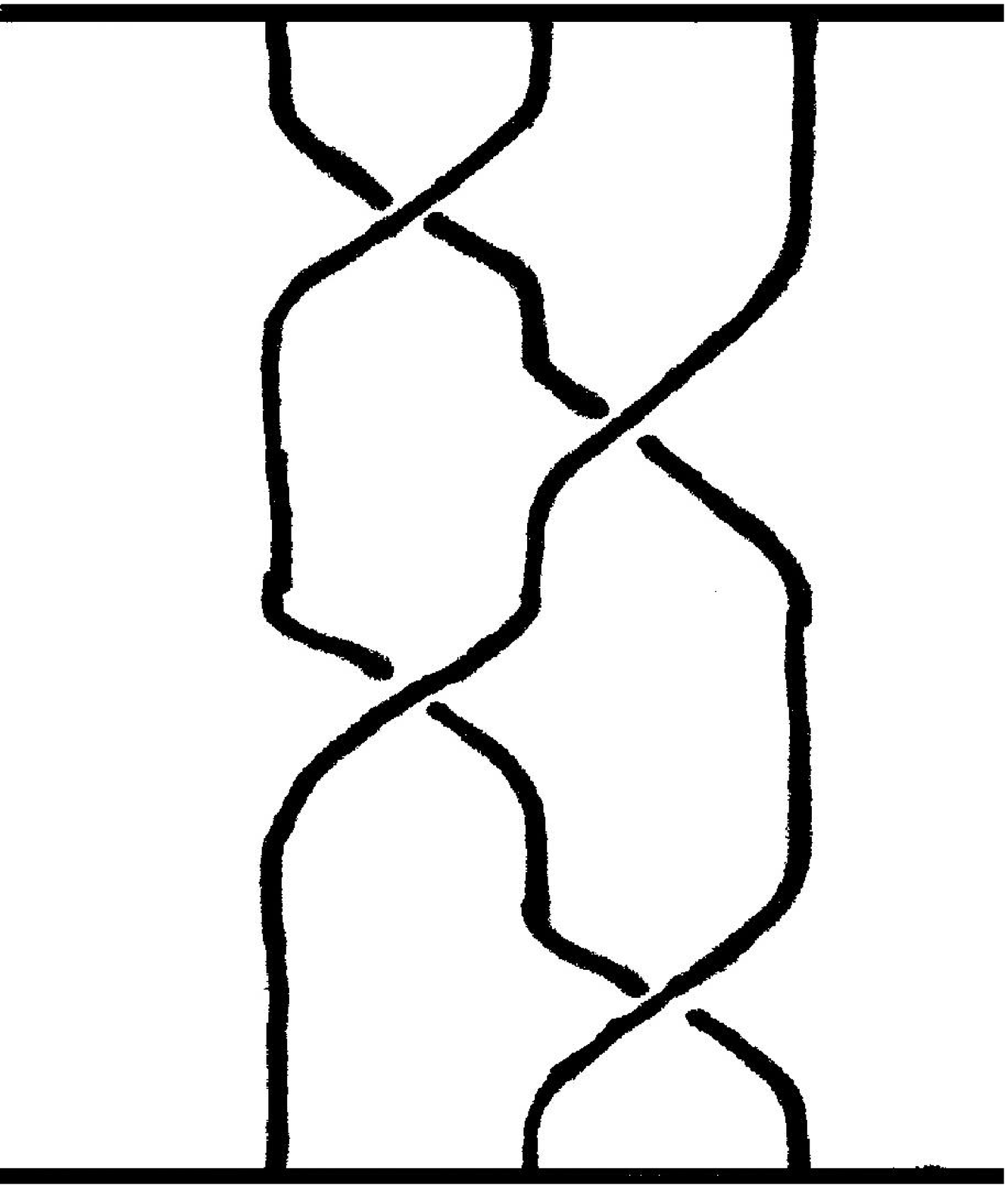}}
\caption{String solution with $k=2/3$, $n=2$ and schematic
three-braid representation of the trefoil knot.}
\end{figure}
Moreover, the observation that all string configurations are
located on a cylinder with constant radius is equivalent to the
fact that the corresponding knots are torus knots. For example,
one can see that our string solutions do not allow us to represent
the figure-eight knot (fig. 3).
%%%%%%%%%%%%%%%%%%%%%%%%%%%%%%%%%%%%%%%%%%%%%%%%%%%%
\subsection{N=2 case}
%%%%%%%%%%%%%%%%%%%%%%%%%%%%%%%%%%%%%%%%%%%%%%%%%%%%
The second simple but interesting case we will investigate, is the
case with two-component solution (\ref{anzatz walc gen}) and with
$c_0=0$. Now one gets
\begin{equation}
2R_1^2R_2^2 \cos[ (k_1-k_2)z +(n_1-n_2)\phi ] + R_1^2+R_2^2 =0.
\label{position solit gen 2}
\end{equation}
Then there is a central string located at the origin
\begin{equation}
\rho =0 \label{pos sol curve2}
\end{equation}
as well as $n_1-n_2$ satellite strings
\begin{equation}
R_1(\rho_0)=R_2(\rho_0), \; \; \; (k_1-k_2)z +(n_1-n_2) \phi =\pi,
+2\pi l \label{pos sol curve3}
\end{equation}
$l=0,1..n_1-n_2$ which lie on the cylinder with a constant radius
$\rho_0$. It can be checked that winding number of the central
string is $Q_c=\mbox{min} \{n_1,n_2\}$ whereas each of the
satellite strings carries the unit topological charge.
%%%%%%%%%%%%%%%%%%%%%%%%%%%%%%%%%%%%%%%%%%%%%%%%%%%%
\subsection{Elliptic configurations}
%%%%%%%%%%%%%%%%%%%%%%%%%%%%%%%%%%%%%%%%%%%%%%%%%%%%
Now, we demonstrate how one can construct a solution describing
strings located on an arbitrary surface topologically equivalent
to cylinder. For example, we adopt before obtained solutions in
the case of the cylindrically-elliptic coordinates $(\eta, \phi,
z)$, where
$$x=\frac{1}{2}a\cos \phi \cosh \eta, $$ $$y=\frac{1}{2}a\sin \phi
\sinh \eta, $$ \begin{equation} z=z. \label{walc} \end{equation}
Then the eikonal equation possesses the following solutions
\begin{equation}
u(\eta, \phi, z)= Ce^{\pm i kz} e^{\pm i\sqrt{\lambda^2
+\frac{k^2a^2}{4}}
E\left(i\eta,1-\frac{4\lambda^2}{4\lambda^2+k^2a^2}\right)}e^{\pm
i \sqrt{\lambda^2 +\frac{k^2a^2}{4}}
E\left(\phi,1-\frac{4\lambda^2}{4\lambda^2+k^2a^2}\right)}+c_0,
\label{eliptic sol}
\end{equation}
where $E$ is elliptic integral the second kind. In order to
guarantee uniqueness of the solution, parameters $\lambda$ and $k$
must fulfill relation
\begin{equation}
\frac{2}{\pi} \sqrt{\lambda^2 +\frac{k^2a^2}{4}} E \left(
1-\frac{4\lambda^2}{4\lambda^2+k^2a^2} \right) =n \in N.
\label{eliptic cond}
\end{equation}
\begin{figure} \center
\resizebox{0.5 \textwidth}{!} {\includegraphics{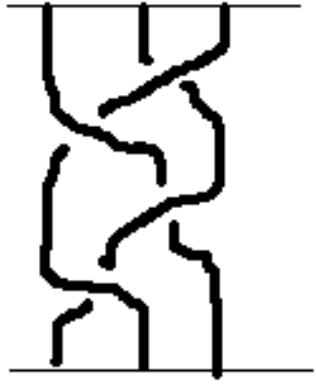}}
\caption{Three-braid representation of the figure-eight knot.}
\end{figure}
This new number $n$ fixes the topological contents of the
configuration. We see that strings lie on a elliptic cylinder with
a constant radius $\eta_0$
\begin{equation}
e^{\pm i\sqrt{\lambda^2 +\frac{k^2a^2}{4}}
E\left(i\eta_0,1-\frac{4\lambda^2}{4\lambda^2+k^2a^2}\right)}=c_0
\label{eliptic radius}
\end{equation}
and are given by the formula
\begin{equation}
kz+\sqrt{\lambda^2 +\frac{k^2a^2}{4}}
E\left(\phi,1-\frac{4\lambda^2}{4\lambda^2+k^2a^2}\right)=\pi
+2\pi l, \label{eliptic string}
\end{equation}
where $l=0,1..n-1$. The generalization to a configuration which
describes strings located on any cylinder (or concentric
cylinders) with an arbitrary shaped base is straightforward. One
has to just introduce the pertinent coordinates and solve the
eikonal equation.
\\ \\
It is also easy to notice that using the simplest topologically
nontrivial solutions $u_0$ (\ref{anzatz walc}), (\ref{eliptic
sol}) with $n=1$ we are able to obtain their multi-string
generalizations (\ref{anzatz walc gen}). Let us for example
consider cylindric string
\begin{equation}
u_0=C  \frac{\rho e^{ \sqrt{k^2\rho^2+1}} e^{ i (\phi
+kz)}}{1+\sqrt{k^2\rho^2+1}} . \label{u0}
\end{equation}
We see that every solution in the form
\begin{equation}
u=F(u_0), \label{sol gen}
\end{equation}
where $F$ is a real and differentiable function, also satisfies
the eikonal equation. In particular, multi-string configurations
are derived by taking $F$ as a polynomial function.
%%%%%%%%%%%%%%%%%%%%%%%%%%%%%%%%%%%%%%%%%%%%%%%%%%%%%%%%%%%%%%%
\section{\bf{Topological Hedgehogs}}
%%%%%%%%%%%%%%%%%%%%%%%%%%%%%%%%%%%%%%%%%%%%%%%%%%%%%%%%%%%%%%%
In this section we take under consideration another type of
topological objects with the non-trivial $\pi_2(S^2)$ homotopy
group i.e. hedgehogs or monopoles. The topology of these
configurations emerges from the index of the map
$\vec{n}_{\infty}: \; S^2 \rightarrow S^2$.
\\
Since the standard hedgehog possesses the rotational symmetry it
is natural to apply the spherical coordinates $(r,\theta,\phi)$.
The eikonal equation can be rewritten as
\begin{equation}
u'^2_r+\frac{1}{r^2}u'^2_{\theta} +\frac{1}{r^2 \sin^2 \theta}
u'^2_{\phi}=0. \label{eikonal sfer1}
\end{equation}
Once again we use the separable variables method and assume the
solution in the form
\begin{equation}
u=f(r) g(\theta) h(\phi). \label{sep sfer}
\end{equation}
Then equation (\ref{eikonal}) reads
\begin{equation}
\frac{f'^2_r}{f^2}+\frac{1}{r^2} \left[ \frac{g'^2_{\theta}}{g^2}
+ \frac{1}{\sin^2 \theta} \frac{h'^2_{\phi}}{h^2} \right]=0.
\label{eikonal sfer2}
\end{equation}
In the standard manner we obtain that function $h(\phi)$ satisfies
\begin{equation}
\frac{h'^2_{\phi}}{h^2} = -n^2 \label{eq sfer phi}
\end{equation}
and have solutions
\begin{equation}
h(\phi) =Ae^{\pm i n \phi}, \label{sol sfer phi}
\end{equation}
where $n \in N$. Analogously the shape function is given by
\begin{equation}
\frac{f'^2_r}{f^2} -\frac{m^2}{r^2}=0. \label{eq sfer r}
\end{equation}
The pertinent solutions are
\begin{equation}
f(r)= B r^{\pm m}, \label{sol sfer r}
\end{equation}
here $m$ is any real number. Now, inserting equations (\ref{eq
sfer phi}) and (\ref{eq sfer r}) we are able to write down
equation for the remaining angular variable
\begin{equation}
\frac{g'^2_{\theta}}{g^2} - \frac{n^2}{\sin^2 \theta} = m^2.
\label{eq sfer theta}
\end{equation}
One can integrate it and find the following solutions
\begin{equation}
g(\theta)=C  \left( \frac{\sqrt{n^2+m^2\sin^2 \theta }-|m|\cos
\theta}{\sqrt{n^2+m^2\sin^2 \theta }+|m|\cos \theta} \right)^{\pm
\frac{|n|}{2}} e^{\mp |m| \arctan \sqrt{\frac{n^2}{m^2} +\tan^2
\theta}}, \label{sol sfer theta}
\end{equation}
parameterize by two, previously introduced numbers $n,m$. Thus,
the general solution of the eikonal equation in the spherical
coordinates reads
$$
u(r,\theta, \phi)= $$
\begin{equation}
Ar^{\pm m} \left( \frac{\sqrt{n^2+m^2\sin^2 \theta }-|m|\cos
\theta}{\sqrt{n^2+m^2\sin^2 \theta }+|m|\cos \theta} \right)^{\pm
\frac{|n|}{2}} e^{\mp |m| \arctan \sqrt{\frac{n^2}{m^2} +\tan^2
\theta}}e^{\pm i n \phi}. \label{sol sfer gen}
\end{equation}
Because of the fact that we are mainly interested in finding of
solutions describing topological hedgehogs the following map must
possess non-zero topological index
\begin{equation}
\vec{n} \; : \; S^2_{(r=\infty,\theta,\phi)} \rightarrow
S^2_{|\vec{n}|=1}. \label{topology}
\end{equation}
In other words, behavior of the unit vector field at the spatial
infinity cannot be trivial. For example all configurations with
$\vec{n}_{\infty}=\overrightarrow{\mbox{const}}.$ are, from our
point of view, uninteresting since they lead to vanishing
topological charge. It results in the fact that any configuration
with the non-zero topological index shoul have $m=0$ in (\ref{sol
sfer r}). Then the function $g$ can be rewritten in simpler form
\begin{equation}
g(\theta)= C \left( \tan \frac{\theta}{2} \right)^{\pm |n|}.
\label{sol sfer theta jez}
\end{equation}
Thus, the general hedgehog configuration takes the form
\begin{equation}
u(\phi,\theta)=Ce^{\pm i n \phi} \left( \tan \frac{\theta}{2}
\right)^{\pm |n|} +c. \label{sol_hegdehog}
\end{equation}
Identically as in the string case this solution can be generalized
to a collection of the hedgehogs
\begin{equation}
u(\phi,\theta)=\sum_{j=1}^N Ce^{\pm i n_j \phi} \left( \tan
\frac{\theta}{2} \right)^{\pm |n_j|} +c, \label{sol_hegdehog1}
\end{equation}
where $N=1,2..$. One can check that the total topological charge,
which tells us how many times the sphere
$S^2_{(r=\infty,\theta,\phi)}$ is winded on the sphere
$S^2_{|\vec{n}|=1}$, is given by the expression
\begin{equation}
Q=\mbox{max} \{ n_j , j=1..N\}. \label{monopol charge}
\end{equation}
Let us now notice that obtained eikonal hedgehog configurations
(\ref{sol_hegdehog1}) can be derived also from a dynamical system
i.e. from $O(3)$ invariant action in the $(3+1)$ dimensional
space-time
\begin{equation}
S=\int d^4x \frac{1}{2} (\partial_{\mu} \vec{n} )^2.
\label{action}
\end{equation}
After expressing the Lagrangian in terms of the complex scalar
field $u$ we find that
\begin{equation}
L= \frac{2 }{(1+|u|^2)^2} \partial_{\mu } u \partial^{\mu } u^*.
\label{O(3)_1}
\end{equation}
Then the pertinent equation of motion reads
\begin{equation}
\frac{\partial_{\mu}\partial^{\mu} u}{(1+|u|^2)^2}
-2(\partial_{\nu} u)^2  \frac{u^*}{(1+|u|^2)^3} =0.
\label{eqmot_O3_1}
\end{equation}
One can check that our configurations (\ref{sol_hegdehog1})
fulfill this equation. It is due to the fact that they solve not
only the eikonal equation but also the Laplace equation
\begin{equation}
\nabla^2 u =0. \label{laplace}
\end{equation}
It should be underline that it is unlikely the string solutions
found in the previous section. In this case no action, from which
strings might be obtained as solutions of the pertinent field
equations, is known.
%%%%%%%%%%%%%%%%%%%%%%%%%%%%%%%%%%%%%%%%%%%%%%%%%%%%%%%%%%%%%%%
\section{\bf{Conclusions}}
%%%%%%%%%%%%%%%%%%%%%%%%%%%%%%%%%%%%%%%%%%%%%%%%%%%%%%%%%%%%%%%
In the present work two kinds of the static topological defects
(strings and hedgehogs) living in three dimensional space have
been discussed. These structures have been investigated by means
of the complex eikonal equation.
\\
In particular, many braided multi-string configurations have been
found. Such multi-string solutions possess well defined
topological charge fixed by the biggest value of the $n_i$
parameter in Ansatz (\ref{anzatz walc gen}), which is nothing else
but the winding number of the vector field around the strings. All
solutions have been presented in the analytical form. The most
important feature of our strings is that they lie on a cylinder
with a constant radius. Since every knot (knotted closed string)
possesses its representation in terms of a braid (knots appear
when we cut off a finite cylinder and identify its bases) we see
that only torus knots (i.e. knots which can be plotted on a torus)
can be constructed. In fact, it was observed in our recent paper.
At this stage it is hard to justify whether problem of deriving of
non-cylindrical strings and non-toroidal knots in the framework of
the eikonal equation is only an artefact of the applied method
(and such solutions should be found), or due to some yet unknown
reasons these solutions are forbidden here. Unfortunately, it is
still an open problem.
\\
On the other hand an important problem associated with a property
of the eikonal knots has been successfully solved. Namely, it has
been proved that the eikonal strings (and in consequence eikonal
knots as well) can be located on any surface topologically
equivalent to a cylinder i.e. a tube with the base given by any
reasonable closed curve, for instance ellipse.
\\
This fact seems to be very important in the context of the
Faddeev-Niemi effective action for the low energy gluodynamics
where particles (glueballs) are described as knotted flux-tubes of
the gauge field. As we discussed it before the properties of the
eikonal knots appear to be quite similar to the properties of the
Faddeev-Niemi hopfions. In particular, the eikonal knots serve as
approximated solutions of the Faddeev-Niemi model. Thus, the
results of our paper might find a physical application if we use
obtained here elliptic string solutions (or other strings located
on a'deformed' cylinder)) and try to construct some new
approximated solutions of the Faddeev-Niemi model.
\\
As we said it before, the problem of the existence of the
non-torus knots is still unsolved. However, there are some hints
indicating that only torus knots can appear in the Faddeev-Niemi
model \cite{govin}. Moreover, no non-torus knot has been observed
using numerical methods \cite{battyde}, \cite{salo}. Undoubtedly,
solving the problem of the existence of non-cylindrical open
strings (or equivalently, non-toroidal closed strings) in the
eikonal equation, we could shed any light on the appearance (or
not) of such structures in the Faddeev-Niemi model. This issue
requires farther studies.
\\ \\
This work is partially supported by Foundation for Polish Science
FNP and ESF "COSLAB" program.

\end{document}